\documentclass[twocolumn,english]{revtex4-2}
\usepackage[T1]{fontenc}
\usepackage[latin9]{inputenc}
\usepackage{color}
\usepackage{babel}
\usepackage{amsmath}
\usepackage{amssymb}
\usepackage{graphicx}
\usepackage[unicode=true,pdfusetitle,
 bookmarks=true,bookmarksnumbered=false,bookmarksopen=false,
 breaklinks=true,pdfborder={0 0 0},pdfborderstyle={},backref=false,colorlinks=true]
 {hyperref}

\makeatletter
\usepackage{braket}

\makeatother

\begin{document}
\title{Robustness of a state with Ising topological order against local projective
measurements}
\author{Sanjeev Kumar}
\affiliation{Department of Theoretical Physics, Tata Institute of Fundamental Research,
Homi Bhabha Road, Navy Nagar, Mumbai 400005, India}
\author{Vikram Tripathi}
\affiliation{Department of Theoretical Physics, Tata Institute of Fundamental Research,
Homi Bhabha Road, Navy Nagar, Mumbai 400005, India}
\date{\today}
\begin{abstract}
We investigate the fragility of a topologically ordered state, namely,
the ground state of a weakly Zeeman perturbed honeycomb Kitaev model to environment induced
decoherence effects mimicked by random local projective measurements.
Our findings show the nonabelian Ising topological order, as quantified
by a tripartite mutual information (the topological entanglement entropy
$\gamma,$) is resilient to such disturbances. Further, $\gamma$
is found to evolve smoothly from a topologically ordered state to
a distribution of trivial states as a function of rate of measurement
(temperature). We assess our model by contrasting it with the Toric
Code limit of the Kitaev model, whose ground state has abelian $Z_{2}$ topological order, and
which has garnered greater attention in the literature of fault-tolerant
quantum computation. The findings reveal the topological order  in the Toric Code limit collapses
rapidly as opposed to our model where it can withstand higher measurement rates.
\end{abstract}
\maketitle

\section{Introduction}

Topologically ordered states are a fascinating class of quantum states
that exhibit exotic properties such as edge modes, anyonic excitations,
and long-range entanglement. These states have the potential to revolutionize
quantum information processing and computing. However, their practical
use may be hindered by the susceptibility of these states to errors
and decoherence. Measurements induce errors in a topologically ordered
state, and understanding the effects of these errors on the state's
properties is essential for the development of robust quantum technologies.
Owing to this, the interplay of projective measurements and topological
order is a topic of current research interest \citep{Vijay_Kitaev,Khemani_Kitaev}.
In this work, we investigate the robustness of topologically ordered
states against local projective measurements. We explore the aftermath
of such measurements on the state's entanglement properties. Our results
provide insights into the mechanisms that enable the resilience of
topologically ordered states against local measurements.

From preparation of a quantum state to its characterization, local
projective measurements are an indispensable tool for tweaking quantum
many-body systems and thereby exploring the rich physics offered by
them \citep{PRXQuantum.3.040337,verresen2021efficiently,PhysRevLett.125.030506,PhysRevResearch.2.033347,PhysRevResearch.2.033255,PhysRevLett.131.200201}.
A well-prepared quantum system is typically prone to decoherence due to
coupling with its environment. A recent
study \citep{PhysRevLett.125.210602} smoothly bridges the impact
of weak continuous disturbance on a quantum many body system with
the discrete measurements being performed on it. Here, we mimic the
environmental impact on a quantum system through spatially-random
local projective measurements on the system. The rate of such random
measurements will thus correspond to the temperature of the environment.

The literature reveals findings of measurement-induced phase transitions
\citep{PhysRevX.9.031009,PhysRevResearch.2.043072,PhysRevB.101.104301,PhysRevLett.125.030505,PhysRevA.104.062405,PhysRevB.99.224307,PhysRevB.100.064204,PhysRevB.101.104302,PhysRevB.98.205136,10.21468/SciPostPhys.15.3.096}
in short-range interacting models. Ergodic systems undergo a transition
from volume-law states to steady states having an area-law entanglement
entropy. Efforts are made to see similar effect in long-range models
as well \citep{PhysRevLett.128.010603,PhysRevLett.128.010604,PhysRevLett.128.010605,PhysRevX.10.041020,vijay2020measurement,PhysRevB.106.224305,PhysRevB.104.094304},
where typically due to quick building of entanglement, the saturating
entanglement entropy is still proportional to the system size. The
effect of local measurements has also been studied for area-law states
in 1D noninteracting fermion hopping models \citep{10.21468/SciPostPhys.14.3.031},
and a measurement induced transition between states of differing ground
state degeneracy has been observed. In this study we are interested
in two dimensional systems where the initial state possesses topological
order. For being able to alter the entanglement structure of steady
states and possible applications of such states in Quantum Information
to engineering dynamical phases of matter, the field has acquired
much attention. We concern ourselves with the ground state of the
2D isotropic Kitaev Hamiltonian \citep{KITAEV20062}. The excitations
are fractionalized quasiparticles -- free Majorana fermions and gapped
half-vortices (visons). More importantly, the short-range model exhibits
$Z_{2}$ topological order \citep{PhysRevLett.96.110404}. One manifestation
of this topological order is that the bipartite von Neumann entanglement
entropy is smaller than the expected area law by a constant amount
$\gamma=\ln2$ known as the topological entanglement entropy (TEE).
Adding a small Zeeman field results in a state with Ising topological
order (ITO) with the same value of $\gamma=\ln2.$ The field also
opens a bulk gap for the Majorana fermions. For fields $B$ exceeding
a few per cent of the Kitaev interaction $J$ (e.g. $B/J\approx0.025$
with ferromagnetic Kitaev interactions and $\mathbf{B}=B(111)$) ,
ITO is lost \citep{PhysRevB.98.014418,PhysRevB.104.245113}. The gap
also provides protection against thermal excitations. We study the
interplay between disentangling measurements and entanglement building
unitary evolution and show it results in steady states with finite
$\gamma.$ We further compare our results with the Toric Code model
which is well-known in the literature of fault-tolerant quantum computation
and whose ground state possess $Z_{2}$ topological order with the
same $\gamma=\ln2$ as ITO. Such a state has recently been prepared
using a specific protocol of projective measurements on real quantum
hardware \citep{satzinger2021realizing}. We find entanglement of
the latter is significantly more fragile against random local projective
measurements.

The remaining sections of the paper are structured as follows: In
Sec. \ref{sec:Model}, we present our model and outline the measurement
protocol. Section \ref{sec:Results} focuses on presenting results,
specifically highlighting the dynamics of TEE. Additionally, this
section delves into the diagnosis of the topologically ordered state
through local projective measurements. Finally, Sec. \ref{sec:Discussions}
provides a summary of our findings and includes a discussion.

\section{Measurement routine \label{sec:Model}}

We consider the following Hamiltonian, 
\begin{equation}
\hat{H}=-\sum_{\left\langle ij\right\rangle _{\eta-\text{{link}}}}\sigma_{i}^{\eta}\sigma_{j}^{\eta}+B\sum_{i,\eta}\sigma_{i}^{\eta},\label{eq:H}
\end{equation}
where the first term on the RHS is an isotropic ferromagnetic Kitaev
interaction on the honeycomb lattice. Here $i$ labels the spin-$1/2$
degrees of freedom on the lattice and $\left\langle ij\right\rangle _{\eta-\text{{link}}}$
denotes the nearest neighbors $i,j$ on the three types of links labeled
$\eta=x,y,z.$ The second term on the RHS is a coupling to an external
Zeeman field in $(1,1,1)$ direction. We set the Kitaev interaction
strength to unity and external magnetic field, $B=0.01,$ which places
the ground state well within the ITO phase. We confirmed that the
small magnetic field preserves the TEE, which we obtained using the
Kitaev-Preskill construction \citep{PhysRevLett.96.110404}. We initialize
the system in the ground state of the Hamiltonian, Eq. (\ref{eq:H}).
The decoherence is introduced via periodically timed local projective
measurements on randomly chosen qubits. The protocol for carrying
out the measurement is described below.

\begin{figure}
\includegraphics[width=1\columnwidth]{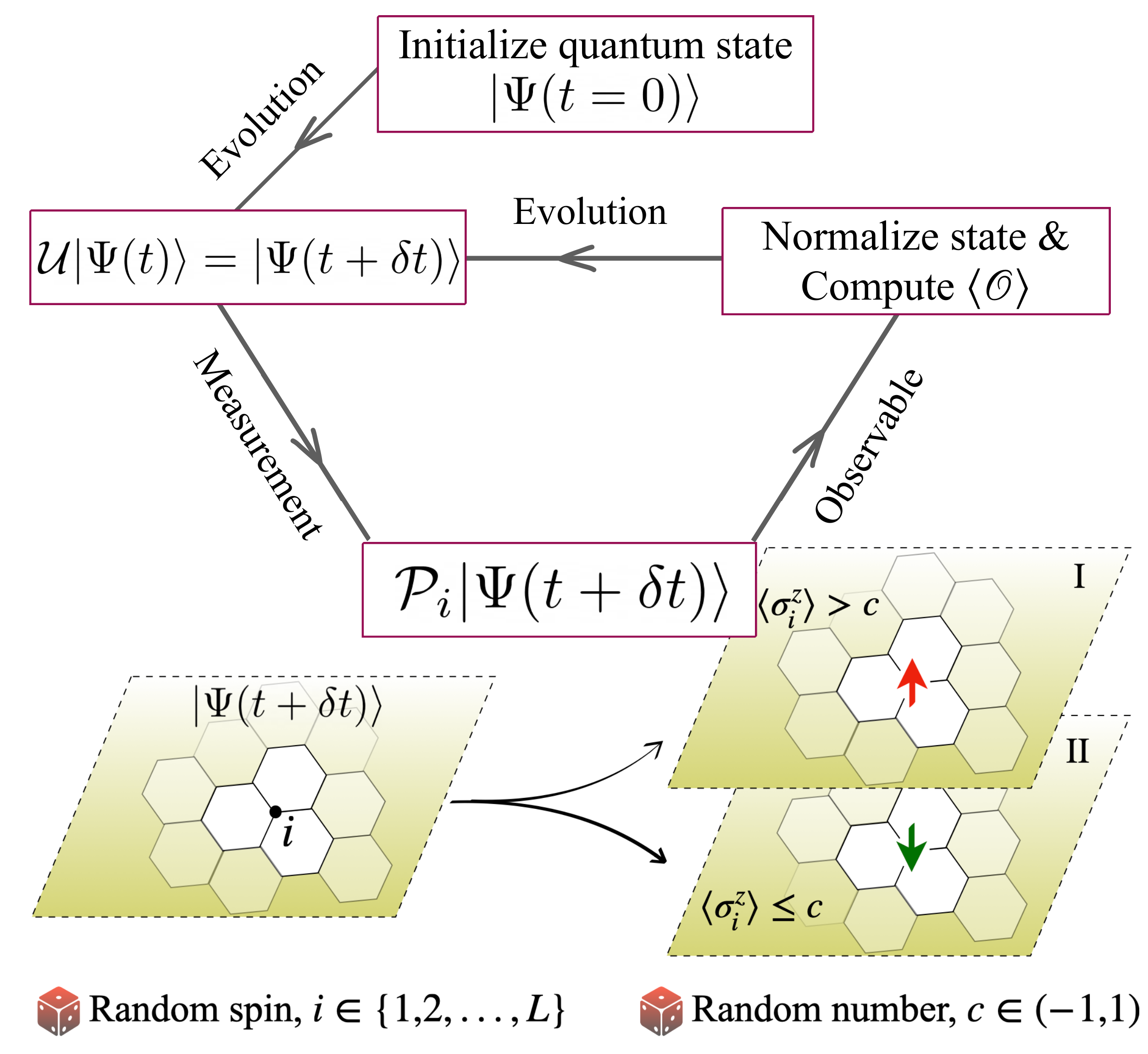}

\caption{\label{fig:Flowchart}Flowchart depicting the measurement protocol:
The system is initialized in a topologically ordered ground state,
$\ket{\psi(0)}$, and unitarily evolved under the Hamiltonian Eq.
(\ref{eq:H}). A local measurement operator, $\mathcal{{P}}_{i}$
selects a spin $i$ from a uniform distribution over the integers
from 1 through $L$, projecting the quantum state onto either of the
shown subspaces depending on the $\braket{\sigma_{i}^{z}}$ and its
comparison with a real number $c$ sampled from a uniform distribution
in $(-1,1)$. The quantum state projection is normalized and observables'
expectation $\braket{\mathcal{{O}}}$ are computed. The observables
are self-averaged over $n\gg1$ runs. Here, $\delta t=\tau/N$ and
$N$ is the number of local projections made periodically (stroboscopically) in the duration
$\tau.$}
\end{figure}

We consider a system undergoing $N$ measurements at equally spaced
intervals over a duration $\tau$. Each measurement involves selecting
a lattice site $i$ from a uniform distribution over ${1,2,...,L}$,
where $L$ denotes the system size. At each chosen site, a local projection
operator acts. We consider the following spin-up and spin-down projection
operators respectively: 
\begin{equation}
\mathcal{{P}^{\pm}}=\frac{1\pm\sigma_{i}^{z}}{2}.
\end{equation}

The projection operator on the $i^{\text{{th}}}-$spin is defined
as 
\begin{equation}
\mathcal{P}_{i}=p\mathcal{{P}^{+}}+(1-p)\mathcal{{P}^{-}},
\end{equation}
where we determine the choice of $p$ (either $0$ or $1$) by comparing
the expectation value of $\sigma_{i}^{z}$ in the instantaneous state
at the time of measurement with a uniformly drawn number $c$ from
the interval $(-1,1).$ For state $\Ket{\Psi}$, satisfying the condition
$\Braket{\Psi|\sigma_{i}^{z}|\Psi}>c$ we impose $p=1$ and project
the state onto the subspace-I where the $i^{\text{{th}}}-$spin is
up. Otherwise, $p=0$ and we project onto the spin-down subspace-II
as shown in Fig. (\ref{fig:Flowchart}). The rationale for introducing
the comparison with a random number is to incorporate the inherent
quantum mechanical probabilities of both possible measurement outcomes
for the state $\ket{\Psi}$, even in single-shot measurements. Simply
determining the outcome based solely on the expectation value $\Braket{\Psi|\sigma_{i}^{z}|\Psi}$
(being greater or less than zero) would not capture the finite probability
of obtaining the less likely outcome. This comparison with $c$, therefore,
serves as a probabilistic mechanism to select the appropriate subspace.
This approach also mimics the actual experimental situation, where
each measurement yields a pure state, which subsequently becomes the
starting point for the next evolution cycle. Following the measurement,
we normalize the obtained quantum state (either I or II) and calculate
expectation values of observables. These expectations are then self-averaged
over a large number of runs $(n\gg1)$. Notably, incorporating a random
number in this manner bears resemblance to the Metropolis algorithm
and presents an alternative strategy for non-unitary stochastic state
evolution, circumventing the need for the density matrix formalism
(see also \citep{PhysRevB.105.094303}). Investigation of decoherence
induced by random local projections has been studied without the random
number approach and specifically using a deterministic criteria wherein
each qubit in a one-dimensional quantum Ising chain model was periodically
probed with a homogeneous probability \citep{PhysRevB.102.035119}.

The unitary evolution attempts to heal the entanglement structure
following a projective measurement. Any particular realization of
$N$ measurements on a system of $L$ qubits (spins) will traverse
one from the total of $n=L^{N}$ possible trajectories. However, keeping
track of exponentially large number of trajectories may not be the
plausible choice; for many observables show self averaging and quickly
reach a steady state value even when averaged over few $(n\gg1)$
typical paths. This in fact is the situation in our simulations. We
keep track of the saturation (i.e. steady state) value dependence
on various measurement rates and duration of the protocols. Our analysis
reveals the emergence of non-equilibrium steady state (NESS) in Kitaev
system. We find a smooth transition from the ground state, which is
a non-trivial state to a trivial state as the measurement rate (temperature)
is increased. However, the exponential decay to the topologically
trivial state is much slower when compared to the anisotropic Kitaev
model,
\begin{equation}
\hat{H}=-\sum_{\left\langle ij\right\rangle _{\eta-\text{{link}}}}J^{\eta}\sigma_{i}^{\eta}\sigma_{j}^{\eta},\label{eq:H-1}
\end{equation}
in the Toric Code limit $|J^{z}|>|J^{x}|+|J^{y}|.$ We explore the
behavior of both the models in the following section.

\section{Results \label{sec:Results}}

As the system is disturbed periodically, we would like to inquire
the manner in which a gapped topologically robust quantum system decoheres.
Signature for the same can be captured in entanglement entropy and
mutual information \citep{PhysRevLett.126.170602,PhysRevB.100.134306,science_Roos,HUR20082208,Wang2017,PhysRevA.91.032313,PhysRevLett.120.040402,bussandri2020sudden}.
We first study the entanglement entropy on account of repeated measurement
on randomly chosen qubits of the system. The Kitaev ground state is
associated with a finite TEE, which is the part of the von Neumann
bipartite entropy, $S_{A}=\text{Tr}\rho_{A}\log\rho_{A}=\alpha L-\gamma$,
remaining after subtracting the area law contribution. Here $L$ is
the perimeter of a 2D subsystem $A$ whose bipartite entanglement
entropy is $S_{A}.$ For the Kitaev ground state, $\gamma=\ln2.$
We numerically obtain $\gamma$ using the well-known Kitaev-Preskill
construction \citep{PhysRevLett.96.110404,PhysRevB.104.245113}. Ground states are obtained through exact diagonalization for a system size of up to $L=24$. The simulation of non-unitary stochastic evolution is conducted using \textsc{quspin} \citep{10.21468/SciPostPhys.2.1.003,10.21468/SciPostPhys.7.2.020}. Note that $S_{A}$, being a non-linear function of $\rho_{A}$ yields two
possibilities for averaging over $n\gg1$ trajectories (measurement
histories), namely $\overline{S_{A}\left(\rho_{A}\right)}$ or $S_{A}\left(\overline{\rho_{A}}\right).$ We consider the former alternative, for it includes solely the quantum
correlations \citep{PhysRevB.105.064305}.

\begin{figure}
\includegraphics[width=0.8\columnwidth]{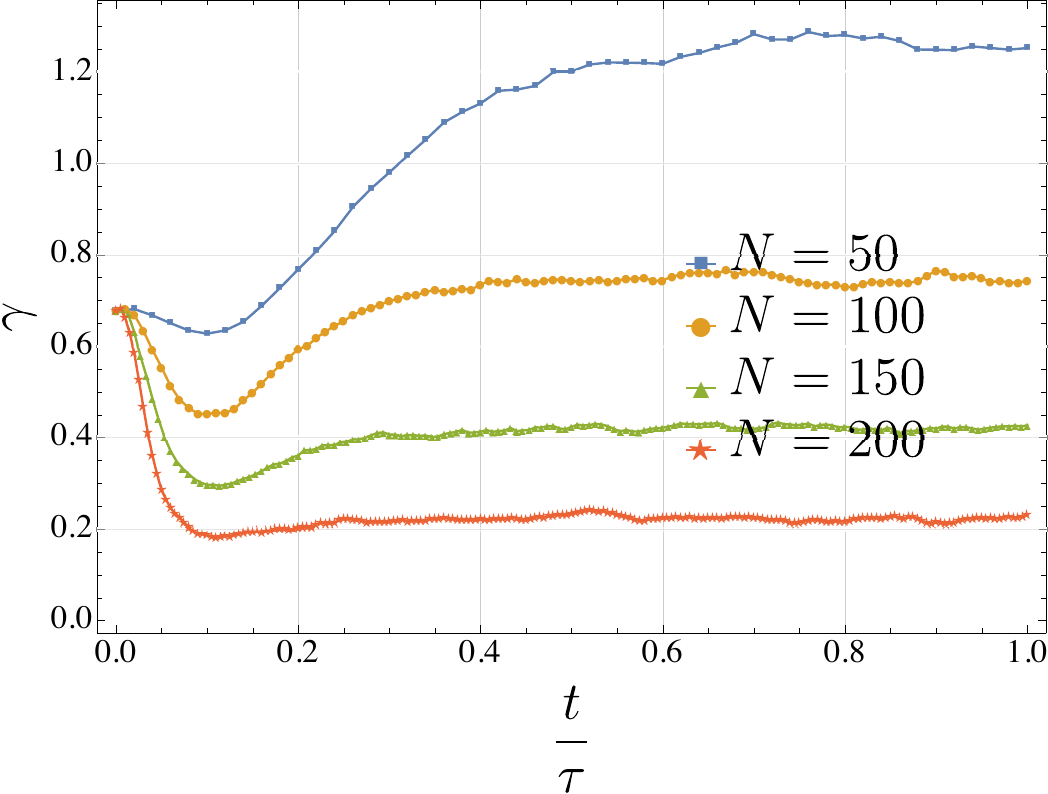}

\caption{\label{fig:Time-dependence}Time evolution of TEE, $\gamma$ for different
rates of measurement. We begin with the ground state of Eq. (\ref{eq:H})
for the system size, $L=16.$ $N$ is the total number of measurements
made periodically in time $\tau=10J^{-1}$. We have set $\hbar=1$.
A total of thousand realizations ($n$) are averaged over to obtain
the plots. The fall and then rise of $\gamma$ at lower $N$ results
from the intricate play of projective measurements and entanglement
building, unitary evolution. For higher $N$, the disruptive measurements
impede the resurgence of mutual information. }
\end{figure}
In Fig. \ref{fig:Time-dependence}, we show the time-dependence of
$\gamma$ for several measurement rates. All the curves begin at the
ground state of Kitaev Hamiltonian in the presence of small Zeeman
field. As periodic measurements are done on the system, the TEE falls.
For lower rates of measurement, the unitary evolution gets enough
time to build-up the entanglement and hence, the system is able to
revive the tripartite mutual information. Saturation values, $\gamma_{\text{{s}}}$
are obtained towards the end of the protocol in Fig. \ref{fig:Time-dependence},
which drop monotonically with the rate of measurement (per site), $r=\cfrac{N}{\tau L}$
. 

\subsection{Non-Equilibrium Steady State}

\begin{figure}
\begin{centering}
\includegraphics[width=0.9\columnwidth]{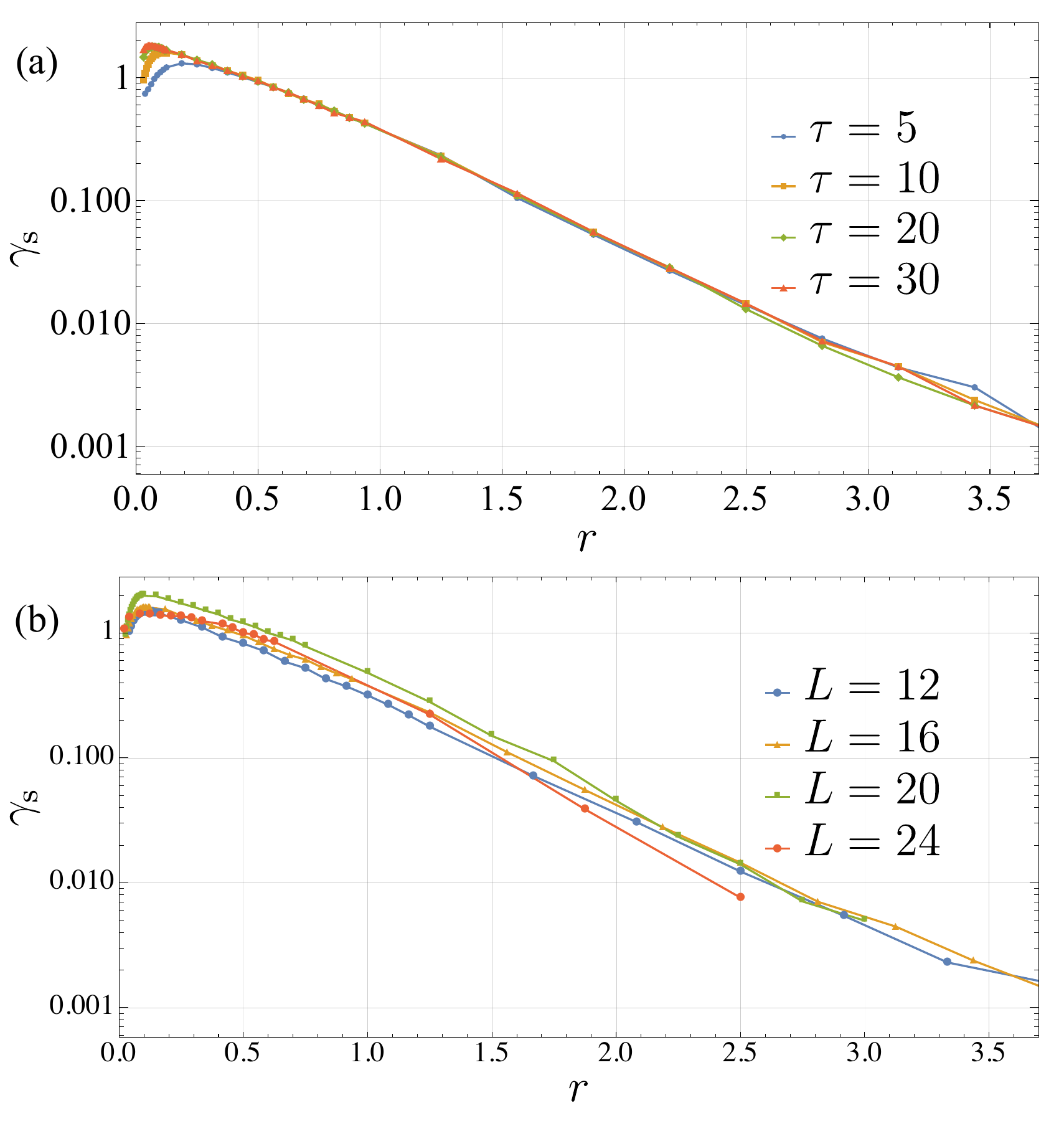}
\par\end{centering}
\caption{\label{fig:NESS}Saturation values of TEE, $\gamma_{\text{{s}}}$
for a Kitaev system subjected to a small Zeeman field $\mathbf{B}=0.01J(111)$
are plotted as obtained after a rate of measurement, $r.$ Overlapping
regions of various time duration $\tau$ protocols in (a) highlight
the establishment of NESS in an  $L=16$-site system. (b) displays $\gamma_{s}$
for different system sizes with $\tau=10J^{-1}$. We see an exponential
decay after a peak. The initial difference in the peaks can be attributed
to the ground state energy gap in various sizes. The decay however,
is indifferent to system sizes.}
\end{figure}
Focusing on Fig. \ref{fig:Time-dependence} we anticipate the establishment
of a NESS. Figure \ref{fig:NESS}, clearly pictures the NESS. At any
given rate of measurement, a steady tripartite mutual information
is established. As shown in Fig. \ref{fig:NESS}, for the Kitaev system
in the ITO phase, the saturation values of TEE follow an exponential
decay after attaining an initial short time peak.

\begin{figure}
\begin{centering}
\includegraphics[width=0.9\columnwidth]{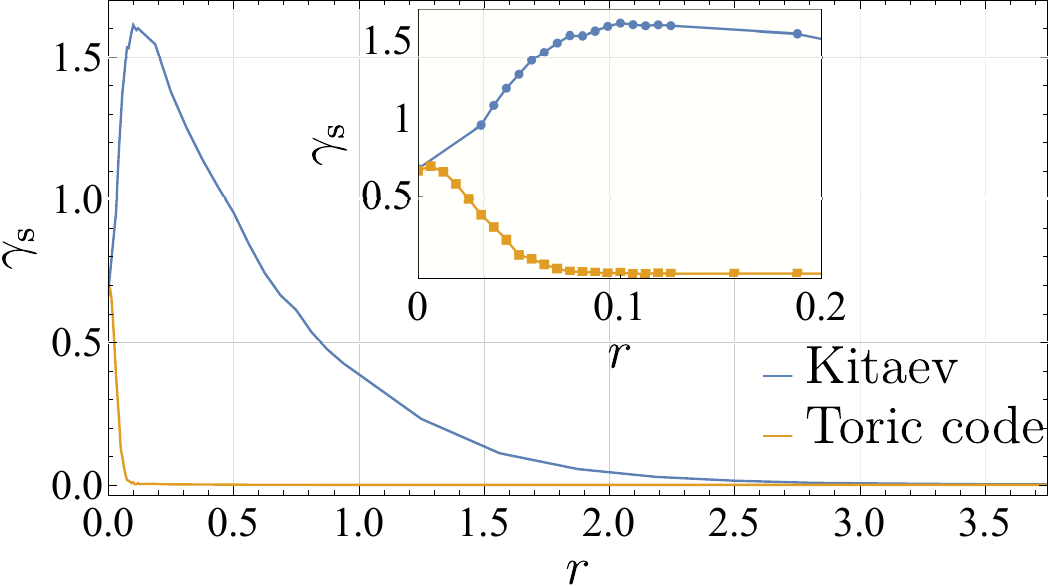}
\par\end{centering}
\caption{\label{fig:Comparison}Comparison of the TEE as withheld by the Kitaev
system, Eq. (\ref{eq:H}) and Toric Code model, Eq. (\ref{eq:H-1})
is shown. Despite the same initial TEE, the latter model shows rapid
decline of the $\gamma_{\text{{s}}}.$ Plots are for the system size
$L=16$ and protocol duration $\tau=10J^{-1}$. For the Toric Code,
$J^{x}=J^{y}=10^{-1}J^{z}$, such that the ground state energies per
site is same for both the models. Inset displays the behavior at small
rates of measurement.}
\end{figure}
Inspection of the saturation values of TEE as a function of rate of
measurement illustrates the relative robustness of $\gamma$ in the
Kitaev system as opposed to the comparable Toric Code model (see Fig.
\ref{fig:Comparison}) which has an abelian $Z_{2}$ topological order
and the same value $\gamma=\ln2$ for the TEE. At small temperatures,
the quantum system is superposed in a state of higher tripartite mutual
information. However, in the latter model, $\gamma$ rapidly collapses
to zero numerically even for a small rate of measurement. When examining
mutual information at nonzero measurement rates, the comparison distinctly
demarcates the two models. The loss of topological order in Toric
Code at any finite temperature has been attributed to vison excitations\citep{grover2020}
in the literature. For comparable energy densities, the Toric Code
limit of the Kitaev model has a smaller vison gap than its Kitaev
counterpart. However the Majorana gap is much smaller in the Kitaev
ITO phase than in the Toric Code phase. This indicates the strong
dependence of the TEE on the gauge sector.

\subsection{Diagnosis of Topological Order}

\begin{figure}
\begin{centering}
\includegraphics[width=0.8\columnwidth]{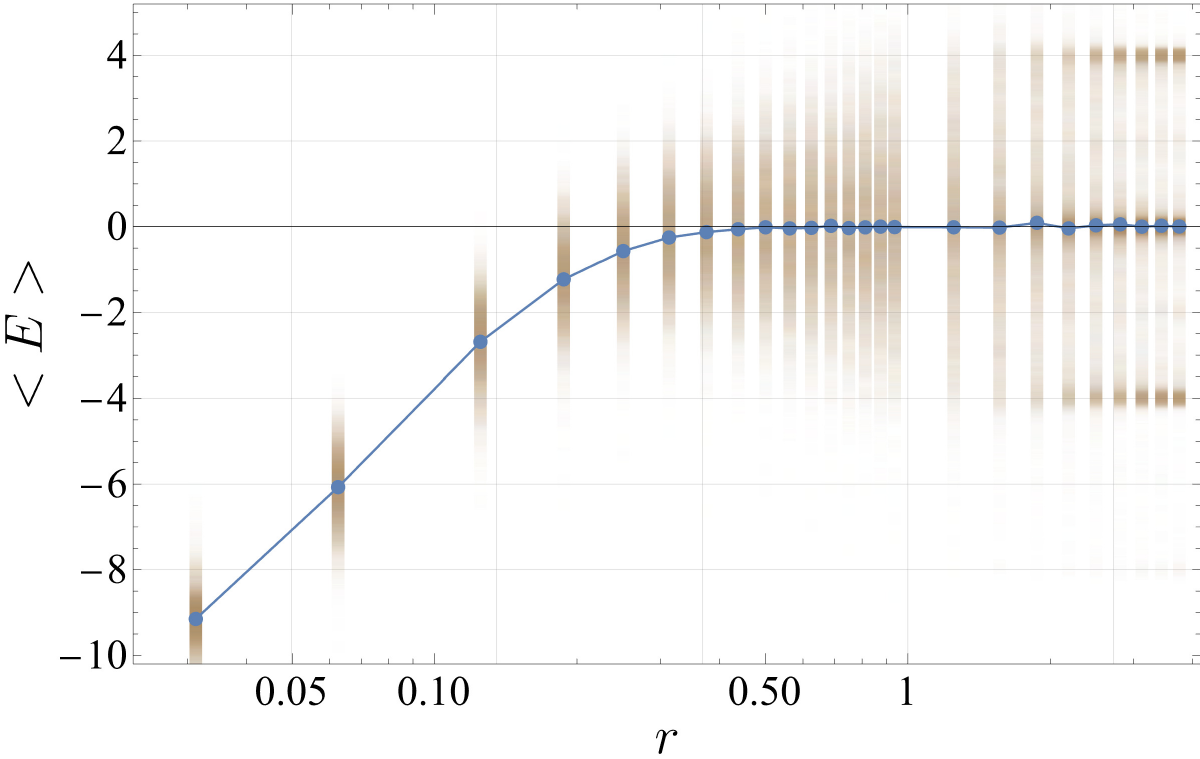}
\par\end{centering}
\caption{\label{fig:Energy}Energy expectation $\braket{E}$ as obtained for
the Kitaev system, Eq. (\ref{eq:H}) with $L=16$ and $\tau=10J^{-1}$
at different measurement rates $r$ is shown. Each blue-colored point
is an average of $n=5000$ brown-colored squares distributed vertically
around it. The latter corresponds to different realizations at $r.$
We see a clear distinction of energy sectors whose superposition result
in the non-vanishing tripartite mutual information. At higher $r$,
this distinction becomes prominent, though, on account of large number
of projective measurements, the state has vanishing $\gamma$ (see
Fig. \ref{fig:Time-dependence}). The entire spectrum lies in the
range $\pm13.4J.$}
\end{figure}
To delve into what exciting superposition the Kitaev system goes through
to yield non-vanishing mutual information, we examine the expectation
values of the Hamiltonian at different rates of random projective
measurements, Fig. \ref{fig:Energy}. Owing to randomness, a given
rate of measurement $r$ in each realization yields, in general, different
energy expectation. Each yellow dot in Fig. \ref{fig:Energy} represents
the energy of the quantum state at the end of the measurement protocol
for an individual run and averaging over these values gives the curve
with the blue dots. Notably, at higher rates of measurement, the energy
distribution has multiple peaks, which is clearly not thermal. The
peak at the centre of the spectrum $E=0$ has approximately $50\%$
weight, and is flanked by two more side peaks ($E=\pm4$ in the figure)
with approximately a quarter weight each. Such structure is not seen
for small $r.$ 

\begin{figure}
\centering{}\includegraphics[width=1\columnwidth]{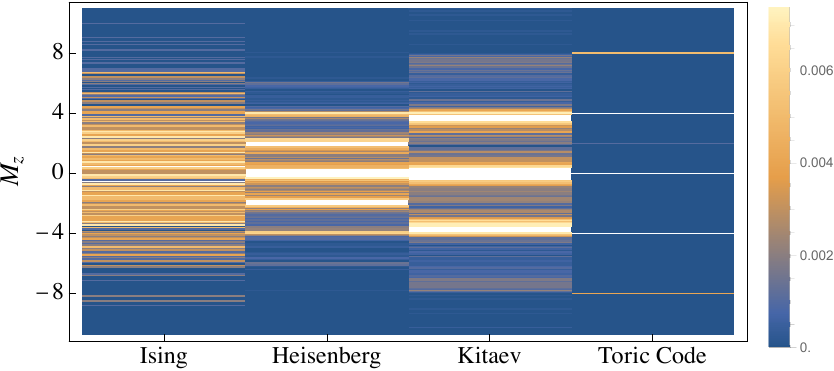}\caption{\label{fig:Diagnosis}Statistical distribution of magnetization $M_{z}=\sum_{i}\sigma_{i}^{z}$
as obtained in different models for the rate of measurement, $r=3.75.$
We have $L=16$ for all the models on a 2D honeycomb lattice. Ising($xx)$,
Heisenberg and Kitaev correspond to the Hamiltonians with coupling
strength of unity in the presence of a small Zeeman field $\mathbf{B}=0.01J(111).$
For the Toric Code limit of the Kitaev model, we have chosen $J^{x}=J^{y}=10^{-1}J^{z}$. In the case of the topologically ordered Kitaev and Toric Code states, distinct bimagnon peaks are seen at $M_z = 0, \pm 4, \ldots$ while in the Ising model a continuous distribution, and in the Heisenberg case, single magnon peaks are seen.}
\end{figure}

Remarkably, even in the high measurement rate regime, signatures of
the initial topologically ordered state continue to persist. Rapid
random projections prohibit state evolution and act as snapshot of
the initial state from different directions in the Hilbert space.
Figure \ref{fig:Diagnosis} illustrates the distribution of total
magnetization $M_{z}=\sum_{i}\sigma_{i}^{z}$ as obtained after the
establishment of the steady state for $r=3.75$ in different models
when starting from their respective ground states. 

We see clear diagnosis of the topologically ordered states, namely,
the ground state of Kitaev and Toric Code model. In these two cases,
the magnetization distribution shows a number of sharp peaks uniformly
separated by steps of two spin flips (bimagnons) around the mean zero
magnetization. In Ising model, we get a continuous distribution while
in Heisenberg model, magnetization separated by a single magnon flip
$(\pm2\text{ units})$ is seen. Owing to the $\sigma^{z}-$projective
measurements, the statistical distribution of $M_{x}$ and $M_{y}$
shows a continuous spread around $0$ magnetization in Kitaev and
Toric Code model.

\section{Discussions\label{sec:Discussions}}

The interplay of entanglement building, unitary evolution and decohering
effect brought about by random projective measurements on the topological
orders of an ITO and $Z_{2}$ ordered state respectively are studied.
The initial and final states in this case are both area-law type and
the transition is in their topological properties. In contrast, in
typical volume-law states, a sharp measurement-induced phase transition
to area-law states as a function of rate of measurement is seen. In
the case of the Kitaev Hamiltonian ground state in the presence of
tiny Zeeman field, we found an exponential decline (with measurement
rate) of the TEE of the final state to a non-ergodic trivial state.
A similar decohering protocol applied to the Toric Code limit of the
Kitaev model reveals that the topological order in the latter is much
more fragile. Earlier studies have suggested the role of vison excitations
in the loss of $Z_{2}$ topological order of the Toric Code at any
non-zero temperature \citep{grover2020}. We found that even at high
rates of projective measurement, the magnetization distribution retains
distinct signatures of the initial topological order in the form of
highly excited bimagnon states around $M_{z}=0.$ These bimagnon peaks
are very reminiscent of quantum scars in nonintegrable systems \citep{PhysRevLett.123.147201,PhysRevB.98.235155,PhysRevB.98.235156,PhysRevB.100.184312}, where ladders of scar states in nonintegrable one-dimensional spin systems have been shown to be generated from any single member of the set through the action of two-particle operators. We believe that states corresponding to our bimagnon peaks do not represent scars because of the projective measurements performed to reach there, and moreover, unlike the topological entanglement entropy of typical Kitaev states which are of the order of $\ln 2,$ here $\gamma$ is measurably much smaller.

\begin{acknowledgments}
The authors acknowledge support of the Department of Atomic Energy,
Government of India, under Project Identification No. RTI 4002, and
the Department of Theoretical Physics, TIFR, for computational resources.
\end{acknowledgments}

\bibliographystyle{apsrev4-2}
\bibliography{ms}

\begin{thebibliography}{50}%
\makeatletter
\providecommand \@ifxundefined [1]{%
 \@ifx{#1\undefined}
}%
\providecommand \@ifnum [1]{%
 \ifnum #1\expandafter \@firstoftwo
 \else \expandafter \@secondoftwo
 \fi
}%
\providecommand \@ifx [1]{%
 \ifx #1\expandafter \@firstoftwo
 \else \expandafter \@secondoftwo
 \fi
}%
\providecommand \natexlab [1]{#1}%
\providecommand \enquote  [1]{``#1''}%
\providecommand \bibnamefont  [1]{#1}%
\providecommand \bibfnamefont [1]{#1}%
\providecommand \citenamefont [1]{#1}%
\providecommand \href@noop [0]{\@secondoftwo}%
\providecommand \href [0]{\begingroup \@sanitize@url \@href}%
\providecommand \@href[1]{\@@startlink{#1}\@@href}%
\providecommand \@@href[1]{\endgroup#1\@@endlink}%
\providecommand \@sanitize@url [0]{\catcode `\\12\catcode `\$12\catcode
  `\&12\catcode `\#12\catcode `\^12\catcode `\_12\catcode `\%12\relax}%
\providecommand \@@startlink[1]{}%
\providecommand \@@endlink[0]{}%
\providecommand \url  [0]{\begingroup\@sanitize@url \@url }%
\providecommand \@url [1]{\endgroup\@href {#1}{\urlprefix }}%
\providecommand \urlprefix  [0]{URL }%
\providecommand \Eprint [0]{\href }%
\providecommand \doibase [0]{https://doi.org/}%
\providecommand \selectlanguage [0]{\@gobble}%
\providecommand \bibinfo  [0]{\@secondoftwo}%
\providecommand \bibfield  [0]{\@secondoftwo}%
\providecommand \translation [1]{[#1]}%
\providecommand \BibitemOpen [0]{}%
\providecommand \bibitemStop [0]{}%
\providecommand \bibitemNoStop [0]{.\EOS\space}%
\providecommand \EOS [0]{\spacefactor3000\relax}%
\providecommand \BibitemShut  [1]{\csname bibitem#1\endcsname}%
\let\auto@bib@innerbib\@empty
\bibitem [{\citenamefont {Lavasani}\ \emph {et~al.}(2023)\citenamefont
  {Lavasani}, \citenamefont {Luo},\ and\ \citenamefont {Vijay}}]{Vijay_Kitaev}%
  \BibitemOpen
  \bibfield  {author} {\bibinfo {author} {\bibfnamefont {A.}~\bibnamefont
  {Lavasani}}, \bibinfo {author} {\bibfnamefont {Z.-X.}\ \bibnamefont {Luo}},\
  and\ \bibinfo {author} {\bibfnamefont {S.}~\bibnamefont {Vijay}},\ }\href
  {https://doi.org/10.1103/PhysRevB.108.115135} {\bibfield  {journal} {\bibinfo
   {journal} {Phys. Rev. B}\ }\textbf {\bibinfo {volume} {108}},\ \bibinfo
  {pages} {115135} (\bibinfo {year} {2023})}\BibitemShut {NoStop}%
\bibitem [{\citenamefont {Sriram}\ \emph {et~al.}(2023)\citenamefont {Sriram},
  \citenamefont {Rakovszky}, \citenamefont {Khemani},\ and\ \citenamefont
  {Ippoliti}}]{Khemani_Kitaev}%
  \BibitemOpen
  \bibfield  {author} {\bibinfo {author} {\bibfnamefont {A.}~\bibnamefont
  {Sriram}}, \bibinfo {author} {\bibfnamefont {T.}~\bibnamefont {Rakovszky}},
  \bibinfo {author} {\bibfnamefont {V.}~\bibnamefont {Khemani}},\ and\ \bibinfo
  {author} {\bibfnamefont {M.}~\bibnamefont {Ippoliti}},\ }\href
  {https://doi.org/10.1103/PhysRevB.108.094304} {\bibfield  {journal} {\bibinfo
   {journal} {Phys. Rev. B}\ }\textbf {\bibinfo {volume} {108}},\ \bibinfo
  {pages} {094304} (\bibinfo {year} {2023})}\BibitemShut {NoStop}%
\bibitem [{\citenamefont {Lu}\ \emph {et~al.}(2022)\citenamefont {Lu},
  \citenamefont {Lessa}, \citenamefont {Kim},\ and\ \citenamefont
  {Hsieh}}]{PRXQuantum.3.040337}%
  \BibitemOpen
  \bibfield  {author} {\bibinfo {author} {\bibfnamefont {T.-C.}\ \bibnamefont
  {Lu}}, \bibinfo {author} {\bibfnamefont {L.~A.}\ \bibnamefont {Lessa}},
  \bibinfo {author} {\bibfnamefont {I.~H.}\ \bibnamefont {Kim}},\ and\ \bibinfo
  {author} {\bibfnamefont {T.~H.}\ \bibnamefont {Hsieh}},\ }\href
  {https://doi.org/10.1103/PRXQuantum.3.040337} {\bibfield  {journal} {\bibinfo
   {journal} {PRX Quantum}\ }\textbf {\bibinfo {volume} {3}},\ \bibinfo {pages}
  {040337} (\bibinfo {year} {2022})}\BibitemShut {NoStop}%
\bibitem [{\citenamefont {Verresen}\ \emph {et~al.}(2021)\citenamefont
  {Verresen}, \citenamefont {Tantivasadakarn},\ and\ \citenamefont
  {Vishwanath}}]{verresen2021efficiently}%
  \BibitemOpen
  \bibfield  {author} {\bibinfo {author} {\bibfnamefont {R.}~\bibnamefont
  {Verresen}}, \bibinfo {author} {\bibfnamefont {N.}~\bibnamefont
  {Tantivasadakarn}},\ and\ \bibinfo {author} {\bibfnamefont {A.}~\bibnamefont
  {Vishwanath}},\ }\href@noop {} {\bibfield  {journal} {\bibinfo  {journal}
  {arXiv preprint arXiv:2112.03061}\ } (\bibinfo {year} {2021})}\BibitemShut
  {NoStop}%
\bibitem [{\citenamefont {Zhang}\ \emph {et~al.}(2020)\citenamefont {Zhang},
  \citenamefont {Zhang}, \citenamefont {Chen}, \citenamefont {Peng},
  \citenamefont {Xu}, \citenamefont {Yin}, \citenamefont {Yu}, \citenamefont
  {Ye}, \citenamefont {Han}, \citenamefont {Xu}, \citenamefont {Chen},
  \citenamefont {Li},\ and\ \citenamefont {Guo}}]{PhysRevLett.125.030506}%
  \BibitemOpen
  \bibfield  {author} {\bibinfo {author} {\bibfnamefont {W.-H.}\ \bibnamefont
  {Zhang}}, \bibinfo {author} {\bibfnamefont {C.}~\bibnamefont {Zhang}},
  \bibinfo {author} {\bibfnamefont {Z.}~\bibnamefont {Chen}}, \bibinfo {author}
  {\bibfnamefont {X.-X.}\ \bibnamefont {Peng}}, \bibinfo {author}
  {\bibfnamefont {X.-Y.}\ \bibnamefont {Xu}}, \bibinfo {author} {\bibfnamefont
  {P.}~\bibnamefont {Yin}}, \bibinfo {author} {\bibfnamefont {S.}~\bibnamefont
  {Yu}}, \bibinfo {author} {\bibfnamefont {X.-J.}\ \bibnamefont {Ye}}, \bibinfo
  {author} {\bibfnamefont {Y.-J.}\ \bibnamefont {Han}}, \bibinfo {author}
  {\bibfnamefont {J.-S.}\ \bibnamefont {Xu}}, \bibinfo {author} {\bibfnamefont
  {G.}~\bibnamefont {Chen}}, \bibinfo {author} {\bibfnamefont {C.-F.}\
  \bibnamefont {Li}},\ and\ \bibinfo {author} {\bibfnamefont {G.-C.}\
  \bibnamefont {Guo}},\ }\href {https://doi.org/10.1103/PhysRevLett.125.030506}
  {\bibfield  {journal} {\bibinfo  {journal} {Phys. Rev. Lett.}\ }\textbf
  {\bibinfo {volume} {125}},\ \bibinfo {pages} {030506} (\bibinfo {year}
  {2020})}\BibitemShut {NoStop}%
\bibitem [{\citenamefont {Roy}\ \emph {et~al.}(2020)\citenamefont {Roy},
  \citenamefont {Chalker}, \citenamefont {Gornyi},\ and\ \citenamefont
  {Gefen}}]{PhysRevResearch.2.033347}%
  \BibitemOpen
  \bibfield  {author} {\bibinfo {author} {\bibfnamefont {S.}~\bibnamefont
  {Roy}}, \bibinfo {author} {\bibfnamefont {J.~T.}\ \bibnamefont {Chalker}},
  \bibinfo {author} {\bibfnamefont {I.~V.}\ \bibnamefont {Gornyi}},\ and\
  \bibinfo {author} {\bibfnamefont {Y.}~\bibnamefont {Gefen}},\ }\href
  {https://doi.org/10.1103/PhysRevResearch.2.033347} {\bibfield  {journal}
  {\bibinfo  {journal} {Phys. Rev. Res.}\ }\textbf {\bibinfo {volume} {2}},\
  \bibinfo {pages} {033347} (\bibinfo {year} {2020})}\BibitemShut {NoStop}%
\bibitem [{\citenamefont {Steiner}\ and\ \citenamefont {von
  Oppen}(2020)}]{PhysRevResearch.2.033255}%
  \BibitemOpen
  \bibfield  {author} {\bibinfo {author} {\bibfnamefont {J.~F.}\ \bibnamefont
  {Steiner}}\ and\ \bibinfo {author} {\bibfnamefont {F.}~\bibnamefont {von
  Oppen}},\ }\href {https://doi.org/10.1103/PhysRevResearch.2.033255}
  {\bibfield  {journal} {\bibinfo  {journal} {Phys. Rev. Res.}\ }\textbf
  {\bibinfo {volume} {2}},\ \bibinfo {pages} {033255} (\bibinfo {year}
  {2020})}\BibitemShut {NoStop}%
\bibitem [{\citenamefont {Zhu}\ \emph {et~al.}(2023)\citenamefont {Zhu},
  \citenamefont {Tantivasadakarn}, \citenamefont {Vishwanath}, \citenamefont
  {Trebst},\ and\ \citenamefont {Verresen}}]{PhysRevLett.131.200201}%
  \BibitemOpen
  \bibfield  {author} {\bibinfo {author} {\bibfnamefont {G.-Y.}\ \bibnamefont
  {Zhu}}, \bibinfo {author} {\bibfnamefont {N.}~\bibnamefont
  {Tantivasadakarn}}, \bibinfo {author} {\bibfnamefont {A.}~\bibnamefont
  {Vishwanath}}, \bibinfo {author} {\bibfnamefont {S.}~\bibnamefont {Trebst}},\
  and\ \bibinfo {author} {\bibfnamefont {R.}~\bibnamefont {Verresen}},\ }\href
  {https://doi.org/10.1103/PhysRevLett.131.200201} {\bibfield  {journal}
  {\bibinfo  {journal} {Phys. Rev. Lett.}\ }\textbf {\bibinfo {volume} {131}},\
  \bibinfo {pages} {200201} (\bibinfo {year} {2023})}\BibitemShut {NoStop}%
\bibitem [{\citenamefont {Szyniszewski}\ \emph {et~al.}(2020)\citenamefont
  {Szyniszewski}, \citenamefont {Romito},\ and\ \citenamefont
  {Schomerus}}]{PhysRevLett.125.210602}%
  \BibitemOpen
  \bibfield  {author} {\bibinfo {author} {\bibfnamefont {M.}~\bibnamefont
  {Szyniszewski}}, \bibinfo {author} {\bibfnamefont {A.}~\bibnamefont
  {Romito}},\ and\ \bibinfo {author} {\bibfnamefont {H.}~\bibnamefont
  {Schomerus}},\ }\href {https://doi.org/10.1103/PhysRevLett.125.210602}
  {\bibfield  {journal} {\bibinfo  {journal} {Phys. Rev. Lett.}\ }\textbf
  {\bibinfo {volume} {125}},\ \bibinfo {pages} {210602} (\bibinfo {year}
  {2020})}\BibitemShut {NoStop}%
\bibitem [{\citenamefont {Skinner}\ \emph {et~al.}(2019)\citenamefont
  {Skinner}, \citenamefont {Ruhman},\ and\ \citenamefont
  {Nahum}}]{PhysRevX.9.031009}%
  \BibitemOpen
  \bibfield  {author} {\bibinfo {author} {\bibfnamefont {B.}~\bibnamefont
  {Skinner}}, \bibinfo {author} {\bibfnamefont {J.}~\bibnamefont {Ruhman}},\
  and\ \bibinfo {author} {\bibfnamefont {A.}~\bibnamefont {Nahum}},\ }\href
  {https://doi.org/10.1103/PhysRevX.9.031009} {\bibfield  {journal} {\bibinfo
  {journal} {Phys. Rev. X}\ }\textbf {\bibinfo {volume} {9}},\ \bibinfo {pages}
  {031009} (\bibinfo {year} {2019})}\BibitemShut {NoStop}%
\bibitem [{\citenamefont {Lunt}\ and\ \citenamefont
  {Pal}(2020)}]{PhysRevResearch.2.043072}%
  \BibitemOpen
  \bibfield  {author} {\bibinfo {author} {\bibfnamefont {O.}~\bibnamefont
  {Lunt}}\ and\ \bibinfo {author} {\bibfnamefont {A.}~\bibnamefont {Pal}},\
  }\href {https://doi.org/10.1103/PhysRevResearch.2.043072} {\bibfield
  {journal} {\bibinfo  {journal} {Phys. Rev. Res.}\ }\textbf {\bibinfo {volume}
  {2}},\ \bibinfo {pages} {043072} (\bibinfo {year} {2020})}\BibitemShut
  {NoStop}%
\bibitem [{\citenamefont {Bao}\ \emph {et~al.}(2020)\citenamefont {Bao},
  \citenamefont {Choi},\ and\ \citenamefont {Altman}}]{PhysRevB.101.104301}%
  \BibitemOpen
  \bibfield  {author} {\bibinfo {author} {\bibfnamefont {Y.}~\bibnamefont
  {Bao}}, \bibinfo {author} {\bibfnamefont {S.}~\bibnamefont {Choi}},\ and\
  \bibinfo {author} {\bibfnamefont {E.}~\bibnamefont {Altman}},\ }\href
  {https://doi.org/10.1103/PhysRevB.101.104301} {\bibfield  {journal} {\bibinfo
   {journal} {Phys. Rev. B}\ }\textbf {\bibinfo {volume} {101}},\ \bibinfo
  {pages} {104301} (\bibinfo {year} {2020})}\BibitemShut {NoStop}%
\bibitem [{\citenamefont {Choi}\ \emph {et~al.}(2020)\citenamefont {Choi},
  \citenamefont {Bao}, \citenamefont {Qi},\ and\ \citenamefont
  {Altman}}]{PhysRevLett.125.030505}%
  \BibitemOpen
  \bibfield  {author} {\bibinfo {author} {\bibfnamefont {S.}~\bibnamefont
  {Choi}}, \bibinfo {author} {\bibfnamefont {Y.}~\bibnamefont {Bao}}, \bibinfo
  {author} {\bibfnamefont {X.-L.}\ \bibnamefont {Qi}},\ and\ \bibinfo {author}
  {\bibfnamefont {E.}~\bibnamefont {Altman}},\ }\href
  {https://doi.org/10.1103/PhysRevLett.125.030505} {\bibfield  {journal}
  {\bibinfo  {journal} {Phys. Rev. Lett.}\ }\textbf {\bibinfo {volume} {125}},\
  \bibinfo {pages} {030505} (\bibinfo {year} {2020})}\BibitemShut {NoStop}%
\bibitem [{\citenamefont {Czischek}\ \emph {et~al.}(2021)\citenamefont
  {Czischek}, \citenamefont {Torlai}, \citenamefont {Ray}, \citenamefont
  {Islam},\ and\ \citenamefont {Melko}}]{PhysRevA.104.062405}%
  \BibitemOpen
  \bibfield  {author} {\bibinfo {author} {\bibfnamefont {S.}~\bibnamefont
  {Czischek}}, \bibinfo {author} {\bibfnamefont {G.}~\bibnamefont {Torlai}},
  \bibinfo {author} {\bibfnamefont {S.}~\bibnamefont {Ray}}, \bibinfo {author}
  {\bibfnamefont {R.}~\bibnamefont {Islam}},\ and\ \bibinfo {author}
  {\bibfnamefont {R.~G.}\ \bibnamefont {Melko}},\ }\href
  {https://doi.org/10.1103/PhysRevA.104.062405} {\bibfield  {journal} {\bibinfo
   {journal} {Phys. Rev. A}\ }\textbf {\bibinfo {volume} {104}},\ \bibinfo
  {pages} {062405} (\bibinfo {year} {2021})}\BibitemShut {NoStop}%
\bibitem [{\citenamefont {Chan}\ \emph {et~al.}(2019)\citenamefont {Chan},
  \citenamefont {Nandkishore}, \citenamefont {Pretko},\ and\ \citenamefont
  {Smith}}]{PhysRevB.99.224307}%
  \BibitemOpen
  \bibfield  {author} {\bibinfo {author} {\bibfnamefont {A.}~\bibnamefont
  {Chan}}, \bibinfo {author} {\bibfnamefont {R.~M.}\ \bibnamefont
  {Nandkishore}}, \bibinfo {author} {\bibfnamefont {M.}~\bibnamefont
  {Pretko}},\ and\ \bibinfo {author} {\bibfnamefont {G.}~\bibnamefont
  {Smith}},\ }\href {https://doi.org/10.1103/PhysRevB.99.224307} {\bibfield
  {journal} {\bibinfo  {journal} {Phys. Rev. B}\ }\textbf {\bibinfo {volume}
  {99}},\ \bibinfo {pages} {224307} (\bibinfo {year} {2019})}\BibitemShut
  {NoStop}%
\bibitem [{\citenamefont {Szyniszewski}\ \emph {et~al.}(2019)\citenamefont
  {Szyniszewski}, \citenamefont {Romito},\ and\ \citenamefont
  {Schomerus}}]{PhysRevB.100.064204}%
  \BibitemOpen
  \bibfield  {author} {\bibinfo {author} {\bibfnamefont {M.}~\bibnamefont
  {Szyniszewski}}, \bibinfo {author} {\bibfnamefont {A.}~\bibnamefont
  {Romito}},\ and\ \bibinfo {author} {\bibfnamefont {H.}~\bibnamefont
  {Schomerus}},\ }\href {https://doi.org/10.1103/PhysRevB.100.064204}
  {\bibfield  {journal} {\bibinfo  {journal} {Phys. Rev. B}\ }\textbf {\bibinfo
  {volume} {100}},\ \bibinfo {pages} {064204} (\bibinfo {year}
  {2019})}\BibitemShut {NoStop}%
\bibitem [{\citenamefont {Jian}\ \emph {et~al.}(2020)\citenamefont {Jian},
  \citenamefont {You}, \citenamefont {Vasseur},\ and\ \citenamefont
  {Ludwig}}]{PhysRevB.101.104302}%
  \BibitemOpen
  \bibfield  {author} {\bibinfo {author} {\bibfnamefont {C.-M.}\ \bibnamefont
  {Jian}}, \bibinfo {author} {\bibfnamefont {Y.-Z.}\ \bibnamefont {You}},
  \bibinfo {author} {\bibfnamefont {R.}~\bibnamefont {Vasseur}},\ and\ \bibinfo
  {author} {\bibfnamefont {A.~W.~W.}\ \bibnamefont {Ludwig}},\ }\href
  {https://doi.org/10.1103/PhysRevB.101.104302} {\bibfield  {journal} {\bibinfo
   {journal} {Phys. Rev. B}\ }\textbf {\bibinfo {volume} {101}},\ \bibinfo
  {pages} {104302} (\bibinfo {year} {2020})}\BibitemShut {NoStop}%
\bibitem [{\citenamefont {Li}\ \emph {et~al.}(2018)\citenamefont {Li},
  \citenamefont {Chen},\ and\ \citenamefont {Fisher}}]{PhysRevB.98.205136}%
  \BibitemOpen
  \bibfield  {author} {\bibinfo {author} {\bibfnamefont {Y.}~\bibnamefont
  {Li}}, \bibinfo {author} {\bibfnamefont {X.}~\bibnamefont {Chen}},\ and\
  \bibinfo {author} {\bibfnamefont {M.~P.~A.}\ \bibnamefont {Fisher}},\ }\href
  {https://doi.org/10.1103/PhysRevB.98.205136} {\bibfield  {journal} {\bibinfo
  {journal} {Phys. Rev. B}\ }\textbf {\bibinfo {volume} {98}},\ \bibinfo
  {pages} {205136} (\bibinfo {year} {2018})}\BibitemShut {NoStop}%
\bibitem [{\citenamefont {Tirrito}\ \emph {et~al.}(2023)\citenamefont
  {Tirrito}, \citenamefont {Santini}, \citenamefont {Fazio},\ and\
  \citenamefont {Collura}}]{10.21468/SciPostPhys.15.3.096}%
  \BibitemOpen
  \bibfield  {author} {\bibinfo {author} {\bibfnamefont {E.}~\bibnamefont
  {Tirrito}}, \bibinfo {author} {\bibfnamefont {A.}~\bibnamefont {Santini}},
  \bibinfo {author} {\bibfnamefont {R.}~\bibnamefont {Fazio}},\ and\ \bibinfo
  {author} {\bibfnamefont {M.}~\bibnamefont {Collura}},\ }\href
  {https://doi.org/10.21468/SciPostPhys.15.3.096} {\bibfield  {journal}
  {\bibinfo  {journal} {SciPost Phys.}\ }\textbf {\bibinfo {volume} {15}},\
  \bibinfo {pages} {096} (\bibinfo {year} {2023})}\BibitemShut {NoStop}%
\bibitem [{\citenamefont {Minato}\ \emph {et~al.}(2022)\citenamefont {Minato},
  \citenamefont {Sugimoto}, \citenamefont {Kuwahara},\ and\ \citenamefont
  {Saito}}]{PhysRevLett.128.010603}%
  \BibitemOpen
  \bibfield  {author} {\bibinfo {author} {\bibfnamefont {T.}~\bibnamefont
  {Minato}}, \bibinfo {author} {\bibfnamefont {K.}~\bibnamefont {Sugimoto}},
  \bibinfo {author} {\bibfnamefont {T.}~\bibnamefont {Kuwahara}},\ and\
  \bibinfo {author} {\bibfnamefont {K.}~\bibnamefont {Saito}},\ }\href
  {https://doi.org/10.1103/PhysRevLett.128.010603} {\bibfield  {journal}
  {\bibinfo  {journal} {Phys. Rev. Lett.}\ }\textbf {\bibinfo {volume} {128}},\
  \bibinfo {pages} {010603} (\bibinfo {year} {2022})}\BibitemShut {NoStop}%
\bibitem [{\citenamefont {Block}\ \emph {et~al.}(2022)\citenamefont {Block},
  \citenamefont {Bao}, \citenamefont {Choi}, \citenamefont {Altman},\ and\
  \citenamefont {Yao}}]{PhysRevLett.128.010604}%
  \BibitemOpen
  \bibfield  {author} {\bibinfo {author} {\bibfnamefont {M.}~\bibnamefont
  {Block}}, \bibinfo {author} {\bibfnamefont {Y.}~\bibnamefont {Bao}}, \bibinfo
  {author} {\bibfnamefont {S.}~\bibnamefont {Choi}}, \bibinfo {author}
  {\bibfnamefont {E.}~\bibnamefont {Altman}},\ and\ \bibinfo {author}
  {\bibfnamefont {N.~Y.}\ \bibnamefont {Yao}},\ }\href
  {https://doi.org/10.1103/PhysRevLett.128.010604} {\bibfield  {journal}
  {\bibinfo  {journal} {Phys. Rev. Lett.}\ }\textbf {\bibinfo {volume} {128}},\
  \bibinfo {pages} {010604} (\bibinfo {year} {2022})}\BibitemShut {NoStop}%
\bibitem [{\citenamefont {M\"uller}\ \emph {et~al.}(2022)\citenamefont
  {M\"uller}, \citenamefont {Diehl},\ and\ \citenamefont
  {Buchhold}}]{PhysRevLett.128.010605}%
  \BibitemOpen
  \bibfield  {author} {\bibinfo {author} {\bibfnamefont {T.}~\bibnamefont
  {M\"uller}}, \bibinfo {author} {\bibfnamefont {S.}~\bibnamefont {Diehl}},\
  and\ \bibinfo {author} {\bibfnamefont {M.}~\bibnamefont {Buchhold}},\ }\href
  {https://doi.org/10.1103/PhysRevLett.128.010605} {\bibfield  {journal}
  {\bibinfo  {journal} {Phys. Rev. Lett.}\ }\textbf {\bibinfo {volume} {128}},\
  \bibinfo {pages} {010605} (\bibinfo {year} {2022})}\BibitemShut {NoStop}%
\bibitem [{\citenamefont {Gullans}\ and\ \citenamefont
  {Huse}(2020)}]{PhysRevX.10.041020}%
  \BibitemOpen
  \bibfield  {author} {\bibinfo {author} {\bibfnamefont {M.~J.}\ \bibnamefont
  {Gullans}}\ and\ \bibinfo {author} {\bibfnamefont {D.~A.}\ \bibnamefont
  {Huse}},\ }\href {https://doi.org/10.1103/PhysRevX.10.041020} {\bibfield
  {journal} {\bibinfo  {journal} {Phys. Rev. X}\ }\textbf {\bibinfo {volume}
  {10}},\ \bibinfo {pages} {041020} (\bibinfo {year} {2020})}\BibitemShut
  {NoStop}%
\bibitem [{\citenamefont {Vijay}(2020)}]{vijay2020measurement}%
  \BibitemOpen
  \bibfield  {author} {\bibinfo {author} {\bibfnamefont {S.}~\bibnamefont
  {Vijay}},\ }\href@noop {} {\bibfield  {journal} {\bibinfo  {journal} {arXiv
  preprint arXiv:2005.03052}\ } (\bibinfo {year} {2020})}\BibitemShut {NoStop}%
\bibitem [{\citenamefont {Sahu}\ \emph {et~al.}(2022)\citenamefont {Sahu},
  \citenamefont {Jian}, \citenamefont {Bentsen},\ and\ \citenamefont
  {Swingle}}]{PhysRevB.106.224305}%
  \BibitemOpen
  \bibfield  {author} {\bibinfo {author} {\bibfnamefont {S.}~\bibnamefont
  {Sahu}}, \bibinfo {author} {\bibfnamefont {S.-K.}\ \bibnamefont {Jian}},
  \bibinfo {author} {\bibfnamefont {G.}~\bibnamefont {Bentsen}},\ and\ \bibinfo
  {author} {\bibfnamefont {B.}~\bibnamefont {Swingle}},\ }\href
  {https://doi.org/10.1103/PhysRevB.106.224305} {\bibfield  {journal} {\bibinfo
   {journal} {Phys. Rev. B}\ }\textbf {\bibinfo {volume} {106}},\ \bibinfo
  {pages} {224305} (\bibinfo {year} {2022})}\BibitemShut {NoStop}%
\bibitem [{\citenamefont {Bentsen}\ \emph {et~al.}(2021)\citenamefont
  {Bentsen}, \citenamefont {Sahu},\ and\ \citenamefont
  {Swingle}}]{PhysRevB.104.094304}%
  \BibitemOpen
  \bibfield  {author} {\bibinfo {author} {\bibfnamefont {G.~S.}\ \bibnamefont
  {Bentsen}}, \bibinfo {author} {\bibfnamefont {S.}~\bibnamefont {Sahu}},\ and\
  \bibinfo {author} {\bibfnamefont {B.}~\bibnamefont {Swingle}},\ }\href
  {https://doi.org/10.1103/PhysRevB.104.094304} {\bibfield  {journal} {\bibinfo
   {journal} {Phys. Rev. B}\ }\textbf {\bibinfo {volume} {104}},\ \bibinfo
  {pages} {094304} (\bibinfo {year} {2021})}\BibitemShut {NoStop}%
\bibitem [{\citenamefont {Kells}\ \emph {et~al.}(2023)\citenamefont {Kells},
  \citenamefont {Meidan},\ and\ \citenamefont
  {Romito}}]{10.21468/SciPostPhys.14.3.031}%
  \BibitemOpen
  \bibfield  {author} {\bibinfo {author} {\bibfnamefont {G.}~\bibnamefont
  {Kells}}, \bibinfo {author} {\bibfnamefont {D.}~\bibnamefont {Meidan}},\ and\
  \bibinfo {author} {\bibfnamefont {A.}~\bibnamefont {Romito}},\ }\href
  {https://doi.org/10.21468/SciPostPhys.14.3.031} {\bibfield  {journal}
  {\bibinfo  {journal} {SciPost Phys.}\ }\textbf {\bibinfo {volume} {14}},\
  \bibinfo {pages} {031} (\bibinfo {year} {2023})}\BibitemShut {NoStop}%
\bibitem [{\citenamefont {Kitaev}(2006)}]{KITAEV20062}%
  \BibitemOpen
  \bibfield  {author} {\bibinfo {author} {\bibfnamefont {A.}~\bibnamefont
  {Kitaev}},\ }\href
  {https://doi.org/https://doi.org/10.1016/j.aop.2005.10.005} {\bibfield
  {journal} {\bibinfo  {journal} {Annals of Physics}\ }\textbf {\bibinfo
  {volume} {321}},\ \bibinfo {pages} {2} (\bibinfo {year} {2006})},\ \bibinfo
  {note} {january Special Issue}\BibitemShut {NoStop}%
\bibitem [{\citenamefont {Kitaev}\ and\ \citenamefont
  {Preskill}(2006)}]{PhysRevLett.96.110404}%
  \BibitemOpen
  \bibfield  {author} {\bibinfo {author} {\bibfnamefont {A.}~\bibnamefont
  {Kitaev}}\ and\ \bibinfo {author} {\bibfnamefont {J.}~\bibnamefont
  {Preskill}},\ }\href {https://doi.org/10.1103/PhysRevLett.96.110404}
  {\bibfield  {journal} {\bibinfo  {journal} {Phys. Rev. Lett.}\ }\textbf
  {\bibinfo {volume} {96}},\ \bibinfo {pages} {110404} (\bibinfo {year}
  {2006})}\BibitemShut {NoStop}%
\bibitem [{\citenamefont {Gohlke}\ \emph {et~al.}(2018)\citenamefont {Gohlke},
  \citenamefont {Moessner},\ and\ \citenamefont
  {Pollmann}}]{PhysRevB.98.014418}%
  \BibitemOpen
  \bibfield  {author} {\bibinfo {author} {\bibfnamefont {M.}~\bibnamefont
  {Gohlke}}, \bibinfo {author} {\bibfnamefont {R.}~\bibnamefont {Moessner}},\
  and\ \bibinfo {author} {\bibfnamefont {F.}~\bibnamefont {Pollmann}},\ }\href
  {https://doi.org/10.1103/PhysRevB.98.014418} {\bibfield  {journal} {\bibinfo
  {journal} {Phys. Rev. B}\ }\textbf {\bibinfo {volume} {98}},\ \bibinfo
  {pages} {014418} (\bibinfo {year} {2018})}\BibitemShut {NoStop}%
\bibitem [{\citenamefont {Kumar}\ \emph {et~al.}(2021)\citenamefont {Kumar},
  \citenamefont {Sharma},\ and\ \citenamefont
  {Tripathi}}]{PhysRevB.104.245113}%
  \BibitemOpen
  \bibfield  {author} {\bibinfo {author} {\bibfnamefont {S.}~\bibnamefont
  {Kumar}}, \bibinfo {author} {\bibfnamefont {S.}~\bibnamefont {Sharma}},\ and\
  \bibinfo {author} {\bibfnamefont {V.}~\bibnamefont {Tripathi}},\ }\href
  {https://doi.org/10.1103/PhysRevB.104.245113} {\bibfield  {journal} {\bibinfo
   {journal} {Phys. Rev. B}\ }\textbf {\bibinfo {volume} {104}},\ \bibinfo
  {pages} {245113} (\bibinfo {year} {2021})}\BibitemShut {NoStop}%
\bibitem [{\citenamefont {Satzinger}\ \emph {et~al.}(2021)\citenamefont
  {Satzinger}, \citenamefont {Liu}, \citenamefont {Smith}, \citenamefont
  {Knapp}, \citenamefont {Newman}, \citenamefont {Jones}, \citenamefont {Chen},
  \citenamefont {Quintana}, \citenamefont {Mi}, \citenamefont {Dunsworth} \emph
  {et~al.}}]{satzinger2021realizing}%
  \BibitemOpen
  \bibfield  {author} {\bibinfo {author} {\bibfnamefont {K.}~\bibnamefont
  {Satzinger}}, \bibinfo {author} {\bibfnamefont {Y.-J.}\ \bibnamefont {Liu}},
  \bibinfo {author} {\bibfnamefont {A.}~\bibnamefont {Smith}}, \bibinfo
  {author} {\bibfnamefont {C.}~\bibnamefont {Knapp}}, \bibinfo {author}
  {\bibfnamefont {M.}~\bibnamefont {Newman}}, \bibinfo {author} {\bibfnamefont
  {C.}~\bibnamefont {Jones}}, \bibinfo {author} {\bibfnamefont
  {Z.}~\bibnamefont {Chen}}, \bibinfo {author} {\bibfnamefont {C.}~\bibnamefont
  {Quintana}}, \bibinfo {author} {\bibfnamefont {X.}~\bibnamefont {Mi}},
  \bibinfo {author} {\bibfnamefont {A.}~\bibnamefont {Dunsworth}}, \emph
  {et~al.},\ }\href {https://doi.org/10.1126/science.abi8378} {\bibfield
  {journal} {\bibinfo  {journal} {Science}\ }\textbf {\bibinfo {volume}
  {374}},\ \bibinfo {pages} {1237} (\bibinfo {year} {2021})}\BibitemShut
  {NoStop}%
\bibitem [{\citenamefont {Coppola}\ \emph {et~al.}(2022)\citenamefont
  {Coppola}, \citenamefont {Tirrito}, \citenamefont {Karevski},\ and\
  \citenamefont {Collura}}]{PhysRevB.105.094303}%
  \BibitemOpen
  \bibfield  {author} {\bibinfo {author} {\bibfnamefont {M.}~\bibnamefont
  {Coppola}}, \bibinfo {author} {\bibfnamefont {E.}~\bibnamefont {Tirrito}},
  \bibinfo {author} {\bibfnamefont {D.}~\bibnamefont {Karevski}},\ and\
  \bibinfo {author} {\bibfnamefont {M.}~\bibnamefont {Collura}},\ }\href
  {https://doi.org/10.1103/PhysRevB.105.094303} {\bibfield  {journal} {\bibinfo
   {journal} {Phys. Rev. B}\ }\textbf {\bibinfo {volume} {105}},\ \bibinfo
  {pages} {094303} (\bibinfo {year} {2022})}\BibitemShut {NoStop}%
\bibitem [{\citenamefont {Rossini}\ and\ \citenamefont
  {Vicari}(2020)}]{PhysRevB.102.035119}%
  \BibitemOpen
  \bibfield  {author} {\bibinfo {author} {\bibfnamefont {D.}~\bibnamefont
  {Rossini}}\ and\ \bibinfo {author} {\bibfnamefont {E.}~\bibnamefont
  {Vicari}},\ }\href {https://doi.org/10.1103/PhysRevB.102.035119} {\bibfield
  {journal} {\bibinfo  {journal} {Phys. Rev. B}\ }\textbf {\bibinfo {volume}
  {102}},\ \bibinfo {pages} {035119} (\bibinfo {year} {2020})}\BibitemShut
  {NoStop}%
\bibitem [{\citenamefont {Alberton}\ \emph {et~al.}(2021)\citenamefont
  {Alberton}, \citenamefont {Buchhold},\ and\ \citenamefont
  {Diehl}}]{PhysRevLett.126.170602}%
  \BibitemOpen
  \bibfield  {author} {\bibinfo {author} {\bibfnamefont {O.}~\bibnamefont
  {Alberton}}, \bibinfo {author} {\bibfnamefont {M.}~\bibnamefont {Buchhold}},\
  and\ \bibinfo {author} {\bibfnamefont {S.}~\bibnamefont {Diehl}},\ }\href
  {https://doi.org/10.1103/PhysRevLett.126.170602} {\bibfield  {journal}
  {\bibinfo  {journal} {Phys. Rev. Lett.}\ }\textbf {\bibinfo {volume} {126}},\
  \bibinfo {pages} {170602} (\bibinfo {year} {2021})}\BibitemShut {NoStop}%
\bibitem [{\citenamefont {Li}\ \emph {et~al.}(2019)\citenamefont {Li},
  \citenamefont {Chen},\ and\ \citenamefont {Fisher}}]{PhysRevB.100.134306}%
  \BibitemOpen
  \bibfield  {author} {\bibinfo {author} {\bibfnamefont {Y.}~\bibnamefont
  {Li}}, \bibinfo {author} {\bibfnamefont {X.}~\bibnamefont {Chen}},\ and\
  \bibinfo {author} {\bibfnamefont {M.~P.~A.}\ \bibnamefont {Fisher}},\ }\href
  {https://doi.org/10.1103/PhysRevB.100.134306} {\bibfield  {journal} {\bibinfo
   {journal} {Phys. Rev. B}\ }\textbf {\bibinfo {volume} {100}},\ \bibinfo
  {pages} {134306} (\bibinfo {year} {2019})}\BibitemShut {NoStop}%
\bibitem [{\citenamefont {Brydges}\ \emph {et~al.}(2019)\citenamefont
  {Brydges}, \citenamefont {Elben}, \citenamefont {Jurcevic}, \citenamefont
  {Vermersch}, \citenamefont {Maier}, \citenamefont {Lanyon}, \citenamefont
  {Zoller}, \citenamefont {Blatt},\ and\ \citenamefont {Roos}}]{science_Roos}%
  \BibitemOpen
  \bibfield  {author} {\bibinfo {author} {\bibfnamefont {T.}~\bibnamefont
  {Brydges}}, \bibinfo {author} {\bibfnamefont {A.}~\bibnamefont {Elben}},
  \bibinfo {author} {\bibfnamefont {P.}~\bibnamefont {Jurcevic}}, \bibinfo
  {author} {\bibfnamefont {B.}~\bibnamefont {Vermersch}}, \bibinfo {author}
  {\bibfnamefont {C.}~\bibnamefont {Maier}}, \bibinfo {author} {\bibfnamefont
  {B.~P.}\ \bibnamefont {Lanyon}}, \bibinfo {author} {\bibfnamefont
  {P.}~\bibnamefont {Zoller}}, \bibinfo {author} {\bibfnamefont
  {R.}~\bibnamefont {Blatt}},\ and\ \bibinfo {author} {\bibfnamefont {C.~F.}\
  \bibnamefont {Roos}},\ }\href {https://doi.org/10.1126/science.aau4963}
  {\bibfield  {journal} {\bibinfo  {journal} {Science}\ }\textbf {\bibinfo
  {volume} {364}},\ \bibinfo {pages} {260} (\bibinfo {year}
  {2019})}\BibitemShut {NoStop}%
\bibitem [{\citenamefont {Hur}(2008)}]{HUR20082208}%
  \BibitemOpen
  \bibfield  {author} {\bibinfo {author} {\bibfnamefont {K.~L.}\ \bibnamefont
  {Hur}},\ }\href {https://doi.org/https://doi.org/10.1016/j.aop.2007.12.003}
  {\bibfield  {journal} {\bibinfo  {journal} {Annals of Physics}\ }\textbf
  {\bibinfo {volume} {323}},\ \bibinfo {pages} {2208} (\bibinfo {year}
  {2008})}\BibitemShut {NoStop}%
\bibitem [{\citenamefont {Wang}\ \emph {et~al.}(2017)\citenamefont {Wang},
  \citenamefont {Yue}, \citenamefont {Yu}, \citenamefont {Gao},\ and\
  \citenamefont {Qin}}]{Wang2017}%
  \BibitemOpen
  \bibfield  {author} {\bibinfo {author} {\bibfnamefont {X.-L.}\ \bibnamefont
  {Wang}}, \bibinfo {author} {\bibfnamefont {Q.-L.}\ \bibnamefont {Yue}},
  \bibinfo {author} {\bibfnamefont {C.-H.}\ \bibnamefont {Yu}}, \bibinfo
  {author} {\bibfnamefont {F.}~\bibnamefont {Gao}},\ and\ \bibinfo {author}
  {\bibfnamefont {S.-J.}\ \bibnamefont {Qin}},\ }\href
  {https://doi.org/10.1038/s41598-017-09332-9} {\bibfield  {journal} {\bibinfo
  {journal} {Scientific Reports}\ }\textbf {\bibinfo {volume} {7}},\ \bibinfo
  {pages} {12122} (\bibinfo {year} {2017})}\BibitemShut {NoStop}%
\bibitem [{\citenamefont {Vald\'es-Hern\'andez}\ \emph
  {et~al.}(2015)\citenamefont {Vald\'es-Hern\'andez}, \citenamefont {Majtey},\
  and\ \citenamefont {Plastino}}]{PhysRevA.91.032313}%
  \BibitemOpen
  \bibfield  {author} {\bibinfo {author} {\bibfnamefont {A.}~\bibnamefont
  {Vald\'es-Hern\'andez}}, \bibinfo {author} {\bibfnamefont {A.~P.}\
  \bibnamefont {Majtey}},\ and\ \bibinfo {author} {\bibfnamefont {A.~R.}\
  \bibnamefont {Plastino}},\ }\href
  {https://doi.org/10.1103/PhysRevA.91.032313} {\bibfield  {journal} {\bibinfo
  {journal} {Phys. Rev. A}\ }\textbf {\bibinfo {volume} {91}},\ \bibinfo
  {pages} {032313} (\bibinfo {year} {2015})}\BibitemShut {NoStop}%
\bibitem [{\citenamefont {G\"arttner}\ \emph {et~al.}(2018)\citenamefont
  {G\"arttner}, \citenamefont {Hauke},\ and\ \citenamefont
  {Rey}}]{PhysRevLett.120.040402}%
  \BibitemOpen
  \bibfield  {author} {\bibinfo {author} {\bibfnamefont {M.}~\bibnamefont
  {G\"arttner}}, \bibinfo {author} {\bibfnamefont {P.}~\bibnamefont {Hauke}},\
  and\ \bibinfo {author} {\bibfnamefont {A.~M.}\ \bibnamefont {Rey}},\ }\href
  {https://doi.org/10.1103/PhysRevLett.120.040402} {\bibfield  {journal}
  {\bibinfo  {journal} {Phys. Rev. Lett.}\ }\textbf {\bibinfo {volume} {120}},\
  \bibinfo {pages} {040402} (\bibinfo {year} {2018})}\BibitemShut {NoStop}%
\bibitem [{\citenamefont {Bussandri}\ \emph {et~al.}(2020)\citenamefont
  {Bussandri}, \citenamefont {Majtey},\ and\ \citenamefont
  {Vald{\'e}s-Hern{\'a}ndez}}]{bussandri2020sudden}%
  \BibitemOpen
  \bibfield  {author} {\bibinfo {author} {\bibfnamefont {D.}~\bibnamefont
  {Bussandri}}, \bibinfo {author} {\bibfnamefont {A.~P.}\ \bibnamefont
  {Majtey}},\ and\ \bibinfo {author} {\bibfnamefont {A.}~\bibnamefont
  {Vald{\'e}s-Hern{\'a}ndez}},\ }\href
  {https://doi.org/10.1088/1751-8121/abaf6e} {\bibfield  {journal} {\bibinfo
  {journal} {Journal of Physics A: Mathematical and Theoretical}\ }\textbf
  {\bibinfo {volume} {53}},\ \bibinfo {pages} {405303} (\bibinfo {year}
  {2020})}\BibitemShut {NoStop}%
\bibitem [{\citenamefont {Weinberg}\ and\ \citenamefont
  {Bukov}(2017)}]{10.21468/SciPostPhys.2.1.003}%
  \BibitemOpen
  \bibfield  {author} {\bibinfo {author} {\bibfnamefont {P.}~\bibnamefont
  {Weinberg}}\ and\ \bibinfo {author} {\bibfnamefont {M.}~\bibnamefont
  {Bukov}},\ }\href {https://doi.org/10.21468/SciPostPhys.2.1.003} {\bibfield
  {journal} {\bibinfo  {journal} {SciPost Phys.}\ }\textbf {\bibinfo {volume}
  {2}},\ \bibinfo {pages} {003} (\bibinfo {year} {2017})}\BibitemShut {NoStop}%
\bibitem [{\citenamefont {Weinberg}\ and\ \citenamefont
  {Bukov}(2019)}]{10.21468/SciPostPhys.7.2.020}%
  \BibitemOpen
  \bibfield  {author} {\bibinfo {author} {\bibfnamefont {P.}~\bibnamefont
  {Weinberg}}\ and\ \bibinfo {author} {\bibfnamefont {M.}~\bibnamefont
  {Bukov}},\ }\href {https://doi.org/10.21468/SciPostPhys.7.2.020} {\bibfield
  {journal} {\bibinfo  {journal} {SciPost Phys.}\ }\textbf {\bibinfo {volume}
  {7}},\ \bibinfo {pages} {020} (\bibinfo {year} {2019})}\BibitemShut {NoStop}%
\bibitem [{\citenamefont {Piccitto}\ \emph {et~al.}(2022)\citenamefont
  {Piccitto}, \citenamefont {Russomanno},\ and\ \citenamefont
  {Rossini}}]{PhysRevB.105.064305}%
  \BibitemOpen
  \bibfield  {author} {\bibinfo {author} {\bibfnamefont {G.}~\bibnamefont
  {Piccitto}}, \bibinfo {author} {\bibfnamefont {A.}~\bibnamefont
  {Russomanno}},\ and\ \bibinfo {author} {\bibfnamefont {D.}~\bibnamefont
  {Rossini}},\ }\href {https://doi.org/10.1103/PhysRevB.105.064305} {\bibfield
  {journal} {\bibinfo  {journal} {Phys. Rev. B}\ }\textbf {\bibinfo {volume}
  {105}},\ \bibinfo {pages} {064305} (\bibinfo {year} {2022})}\BibitemShut
  {NoStop}%
\bibitem [{\citenamefont {Lu}\ \emph {et~al.}(2020)\citenamefont {Lu},
  \citenamefont {Hsieh},\ and\ \citenamefont {Grover}}]{grover2020}%
  \BibitemOpen
  \bibfield  {author} {\bibinfo {author} {\bibfnamefont {T.-C.}\ \bibnamefont
  {Lu}}, \bibinfo {author} {\bibfnamefont {T.~H.}\ \bibnamefont {Hsieh}},\ and\
  \bibinfo {author} {\bibfnamefont {T.}~\bibnamefont {Grover}},\ }\href
  {https://doi.org/10.1103/PhysRevLett.125.116801} {\bibfield  {journal}
  {\bibinfo  {journal} {Phys. Rev. Lett.}\ }\textbf {\bibinfo {volume} {125}},\
  \bibinfo {pages} {116801} (\bibinfo {year} {2020})}\BibitemShut {NoStop}%
\bibitem [{\citenamefont {Schecter}\ and\ \citenamefont
  {Iadecola}(2019)}]{PhysRevLett.123.147201}%
  \BibitemOpen
  \bibfield  {author} {\bibinfo {author} {\bibfnamefont {M.}~\bibnamefont
  {Schecter}}\ and\ \bibinfo {author} {\bibfnamefont {T.}~\bibnamefont
  {Iadecola}},\ }\href {https://doi.org/10.1103/PhysRevLett.123.147201}
  {\bibfield  {journal} {\bibinfo  {journal} {Phys. Rev. Lett.}\ }\textbf
  {\bibinfo {volume} {123}},\ \bibinfo {pages} {147201} (\bibinfo {year}
  {2019})}\BibitemShut {NoStop}%
\bibitem [{\citenamefont {Moudgalya}\ \emph
  {et~al.}(2018{\natexlab{a}})\citenamefont {Moudgalya}, \citenamefont
  {Rachel}, \citenamefont {Bernevig},\ and\ \citenamefont
  {Regnault}}]{PhysRevB.98.235155}%
  \BibitemOpen
  \bibfield  {author} {\bibinfo {author} {\bibfnamefont {S.}~\bibnamefont
  {Moudgalya}}, \bibinfo {author} {\bibfnamefont {S.}~\bibnamefont {Rachel}},
  \bibinfo {author} {\bibfnamefont {B.~A.}\ \bibnamefont {Bernevig}},\ and\
  \bibinfo {author} {\bibfnamefont {N.}~\bibnamefont {Regnault}},\ }\href
  {https://doi.org/10.1103/PhysRevB.98.235155} {\bibfield  {journal} {\bibinfo
  {journal} {Phys. Rev. B}\ }\textbf {\bibinfo {volume} {98}},\ \bibinfo
  {pages} {235155} (\bibinfo {year} {2018}{\natexlab{a}})}\BibitemShut
  {NoStop}%
\bibitem [{\citenamefont {Moudgalya}\ \emph
  {et~al.}(2018{\natexlab{b}})\citenamefont {Moudgalya}, \citenamefont
  {Regnault},\ and\ \citenamefont {Bernevig}}]{PhysRevB.98.235156}%
  \BibitemOpen
  \bibfield  {author} {\bibinfo {author} {\bibfnamefont {S.}~\bibnamefont
  {Moudgalya}}, \bibinfo {author} {\bibfnamefont {N.}~\bibnamefont
  {Regnault}},\ and\ \bibinfo {author} {\bibfnamefont {B.~A.}\ \bibnamefont
  {Bernevig}},\ }\href {https://doi.org/10.1103/PhysRevB.98.235156} {\bibfield
  {journal} {\bibinfo  {journal} {Phys. Rev. B}\ }\textbf {\bibinfo {volume}
  {98}},\ \bibinfo {pages} {235156} (\bibinfo {year}
  {2018}{\natexlab{b}})}\BibitemShut {NoStop}%
\bibitem [{\citenamefont {Iadecola}\ \emph {et~al.}(2019)\citenamefont
  {Iadecola}, \citenamefont {Schecter},\ and\ \citenamefont
  {Xu}}]{PhysRevB.100.184312}%
  \BibitemOpen
  \bibfield  {author} {\bibinfo {author} {\bibfnamefont {T.}~\bibnamefont
  {Iadecola}}, \bibinfo {author} {\bibfnamefont {M.}~\bibnamefont {Schecter}},\
  and\ \bibinfo {author} {\bibfnamefont {S.}~\bibnamefont {Xu}},\ }\href
  {https://doi.org/10.1103/PhysRevB.100.184312} {\bibfield  {journal} {\bibinfo
   {journal} {Phys. Rev. B}\ }\textbf {\bibinfo {volume} {100}},\ \bibinfo
  {pages} {184312} (\bibinfo {year} {2019})}\BibitemShut {NoStop}%
\end{thebibliography}%

\end{document}